\documentclass[slac_one]{revtex4}
\usepackage{graphicx}
\usepackage{fancyhdr}
\usepackage{amsmath}
\usepackage{amssymb}
\usepackage{epsfig}
\def\rQCED{{\rm QCED}}

\newcommand {\pslash}{\hbox{$\not\hbox{\kern-2.3pt $p$}$}}

\begin{document}
\title{Precision QED$\otimes$QCD Resummation Theory for LHC Physics:  IR-Improved Scheme for Parton Distributions, Kernels, Reduced Cross Sections with Shower/ME Matching} 

%

\author{B.F.L. Ward, S, Joseph, S. Majhi}
\affiliation{Department of Physics, Baylor University, Waco, TX 76798-7316, USA}
\author{S.A. Yost}
\affiliation{Department of Physics, The Citadel, Charleston, SC 29409, USA}

\begin{abstract}
We summarize the recent progress in a new approach to precision
LHC physics based on
the IR-improved DGLAP-CS theory as it relates to 
a new MC friendly exponentiated scheme for precision calculation of higher 
order corrections to LHC physics in which IR singularities from both QED and 
QCD are canceled to all orders in $\alpha$ and in $\alpha_s$ 
simultaneously in the 
presence of rigorous shower/ME matching. 
We present the first MC data 
comparing the implied new showers themselves with the standard ones using 
the HERWIG6.5 MC event generator as a test case at LHC energies.
\end{abstract}
\maketitle

\thispagestyle{fancy}

The advent of the LHC marks the era of precision QCD, by which we mean
predictions for QCD processes at the total theoretical precision 
tag of $1\%$ or better. 
Here, we summarize the elements
of our approach to such precision calculations~\cite{qced,irdglap1,irdglap2} and its recent applications in Monte Carlo(MC) event generator studies,
which are still preliminary.\par
At such a precision as we have as our goal, issues such as the role of QED~\cite{qedeffects,radcor-ew} are an integral part of the discussion 
and we deal with this by the simultaneous
resummation of QED and QCD large infrared(IR) 
effects, $QED\otimes QCD$ resummation
~\cite{qced} in the presence of parton showers, to be realized on an 
event-by-event basis by MC methods. In proceeding with our
analytical discussion, we first review
this approach to resummation and its relationship to those in Refs.~\cite{cattrent,scet}; this review is 
followed by a summary of the attendant new IR-improved DGLAP-CS theory.
We conclude with a summary of illustrative applications and results,
including the sample MC data.
\par
To put the discussion in the proper perspective, we note that the authors in
Ref.~\cite{scott1} have argued that
the current state-of-the-art theoretical precision tag on single Z
production at the LHC is $(4.1\pm0.3)\%=(1.51\pm 0.75)\%(QCD)\oplus 3.79(PDF)\oplus 0.38\pm 0.26(EW)\%$.
\footnote{Recently, the 
analogous estimate for single W production has been given ~\cite{scott2} as $\sim 5.7$\%.}\par
Turning now to our analysis, we note that, in Refs.~\cite{qced,irdglap1,irdglap2}, we have derived the following expression for the 
hard cross sections in the SM $SU_{2L}\times U_1\times SU_3^c$ EW-QCD theory
\begin{eqnarray}
d\hat\sigma_{\rm exp} = e^{\rm SUM_{IR}(QCED)}
   \sum_{{n,m}=0}^\infty\frac{1}{n!m!}\int\prod_{j_1=1}^n\frac{d^3k_{j_1}}{k_{j_1}} \cr
\prod_{j_2=1}^m\frac{d^3{k'}_{j_2}}{{k'}_{j_2}}
\int\frac{d^4y}{(2\pi)^4}e^{iy\cdot(p_1+q_1-p_2-q_2-\sum k_{j_1}-\sum {k'}_{j_2})+
D_\rQCED} \cr
\tilde{\bar\beta}_{n,m}(k_1,\ldots,k_n;k'_1,\ldots,k'_m)\frac{d^3p_2}{p_2^{\,0}}\frac{d^3q_2}{q_2^{\,0}},
\label{subp15b}
\end{eqnarray}
where the new YFS-style~\cite{yfs} residuals
$\tilde{\bar\beta}_{n,m}(k_1,\ldots,k_n;k'_1,\ldots,k'_m)$ have $n$ hard gluons and $m$ hard photons and we show the final state with two hard final
partons with momenta $p_2,\; q_2$ specified for a generic 2f final state for
definiteness. The infrared functions ${\rm SUM_{IR}(QCED)},\; D_\rQCED\; $
are defined in Refs.~\cite{qced,irdglap1,irdglap2}. This is the simultaneous resummation of QED and QCD large IR effects. Eq.(\ref{subp15b}) is exact.\par
Our approach to QCD resummation is fully consistent with that of
Refs.~\cite{cattrent,scet} as follows. First, Ref.~\cite{geor1} has shown that the latter two approaches are equivalent. We show in Refs.~\cite{irdglap1,irdglap2}
that our approach is consistent with that of Refs.~\cite{cattrent}
by exhibiting the transformation prescription from the resummation formula
for the theory in Refs.~\cite{cattrent} for the generic $2\rightarrow n$ parton process as given in Ref.~\cite{madg} to our theory as given for QCD by restricting (\ref{subp15b}) to its QCD component.\par
We show in Refs.~\cite{irdglap1,irdglap2} that the result (\ref{subp15b})
allows us to improve in the IR regime \footnote{This 
should be distinguished from the also important
resummation in parton density evolution for the $z\rightarrow 0$ regime,
where Regge asymptotics obtain -- see for example Ref.~\cite{ermlv,guido}. This
improvement must also be taken into account for precision LHC predictions.} 
the kernels in DGLAP-CS~\cite{dglap,cs}
theory as follows, using a standard notation:
\begin{align}
P_{qq}(z)&= C_F F_{YFS}(\gamma_q)e^{\frac{1}{2}\delta_q}\left[\frac{1+z^2}{1-z}(1-z)^{\gamma_q} -f_q(\gamma_q)\delta(1-z)\right],\nonumber\\
P_{Gq}(z)&= C_F F_{YFS}(\gamma_q)e^{\frac{1}{2}\delta_q}\frac{1+(1-z)^2}{z} z^{\gamma_q},\nonumber\\
P_{GG}(z)&= 2C_G F_{YFS}(\gamma_G)e^{\frac{1}{2}\delta_G}\{ \frac{1-z}{z}z^{\gamma_G}+\frac{z}{1-z}(1-z)^{\gamma_G}\nonumber\\
&\qquad +\frac{1}{2}(z^{1+\gamma_G}(1-z)+z(1-z)^{1+\gamma_G}) - f_G(\gamma_G) \delta(1-z)\},\nonumber\\
P_{qG}(z)&= F_{YFS}(\gamma_G)e^{\frac{1}{2}\delta_G}\frac{1}{2}\{ z^2(1-z)^{\gamma_G}+(1-z)^2z^{\gamma_G}\}.
\label{dglap19}
\end{align}
These results are being implemented by MC methods as we exhibit already 
below. We turn next to illustrative results and implications.\par
\begin{figure*}[ht]
\centering
\setlength{\unitlength}{0.1mm}
\begin{picture}(1000, 962.5)
\put( 281.25, 956.25){\makebox(0,0)[cb]{\bf (a)} }
\put(768.75, 956.25){\makebox(0,0)[cb]{\bf (b)} }
\put(   0, 543.75){\makebox(0,0)[lb]{\epsfig{file=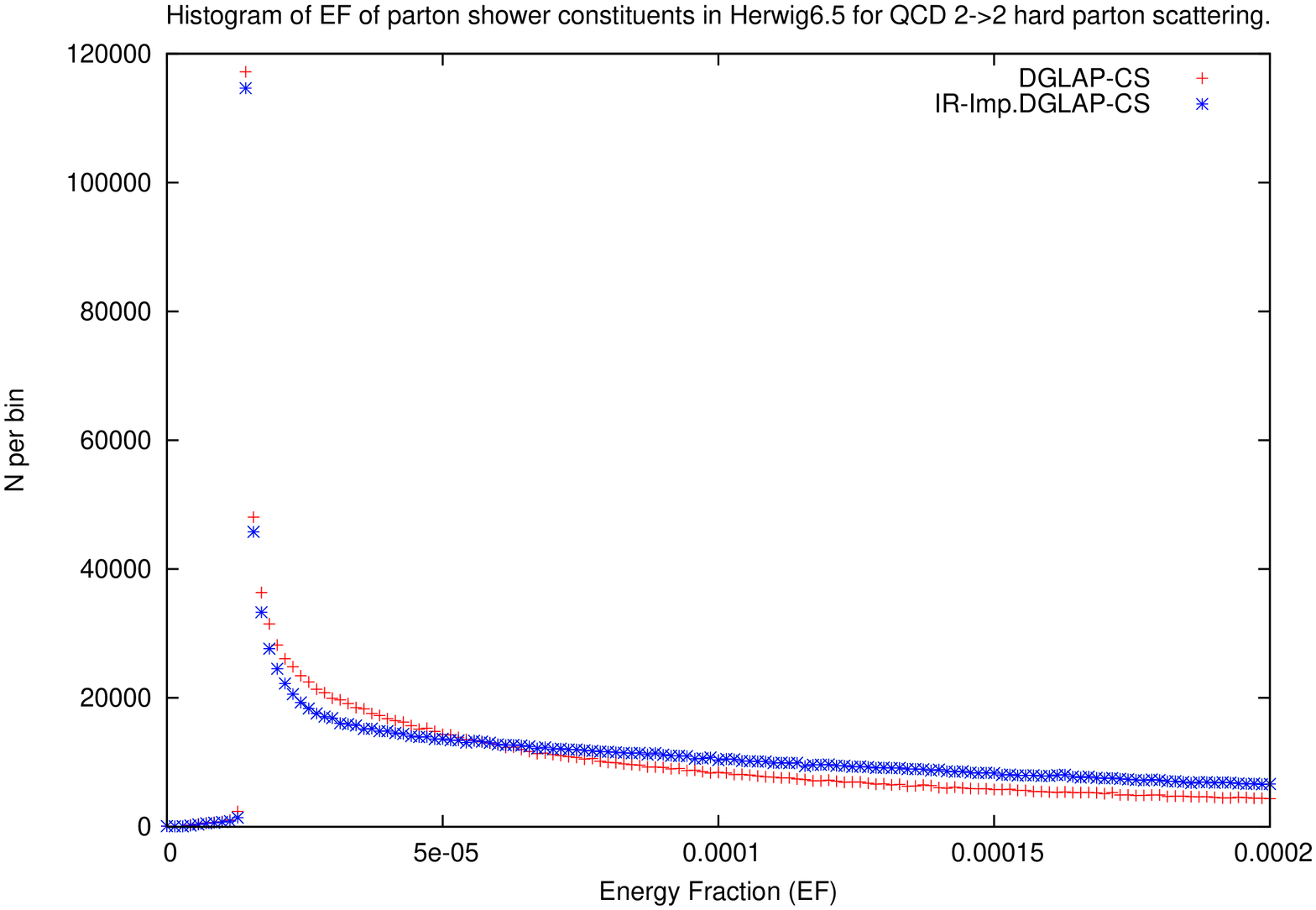,angle=270,
                                        width=50mm}}}
\put( 500, 543.75){\makebox(0,0)[lb]{\epsfig{file=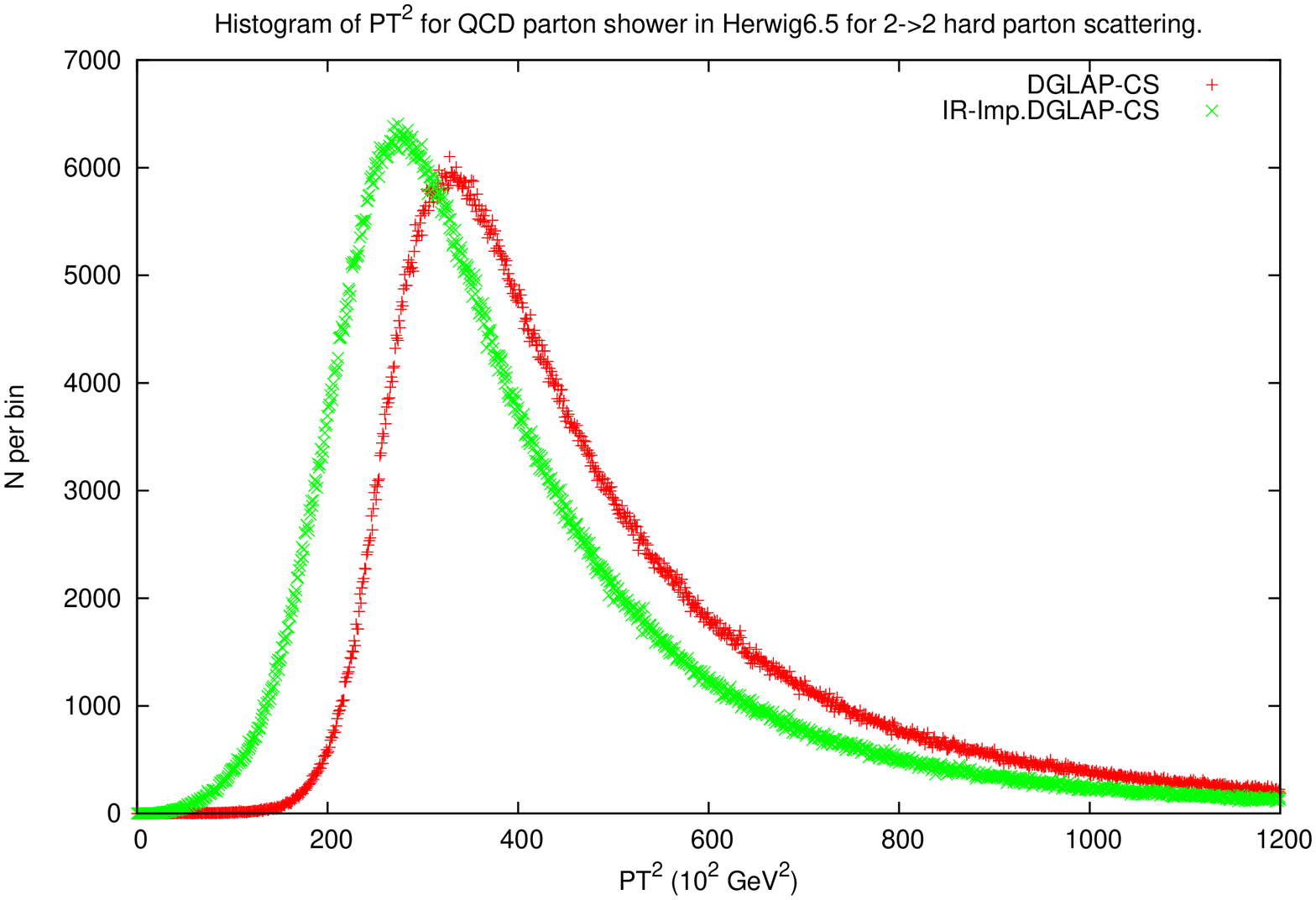,angle=270,
                                        width=50mm}}}
\put( 281.25, 412.5){\makebox(0,0)[cb]{\bf (c)} }
\put(768.75, 412.5){\makebox(0,0)[cb]{\bf (d)} }
\put(   0, 0){\makebox(0,0)[lb]{\epsfig{file=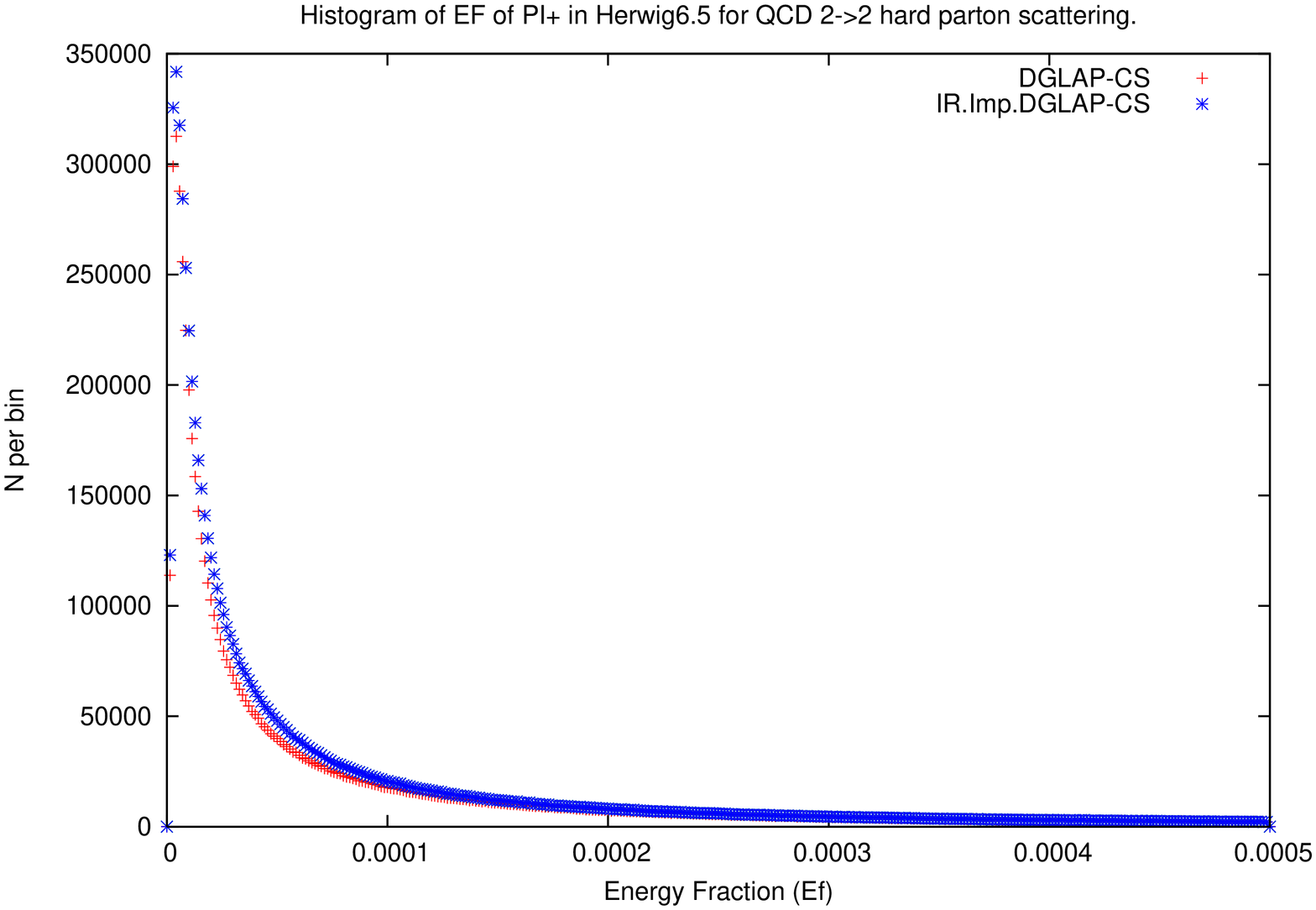,angle=270,
                                        width=50mm}}}
\put( 500, 0){\makebox(0,0)[lb]{\epsfig{file=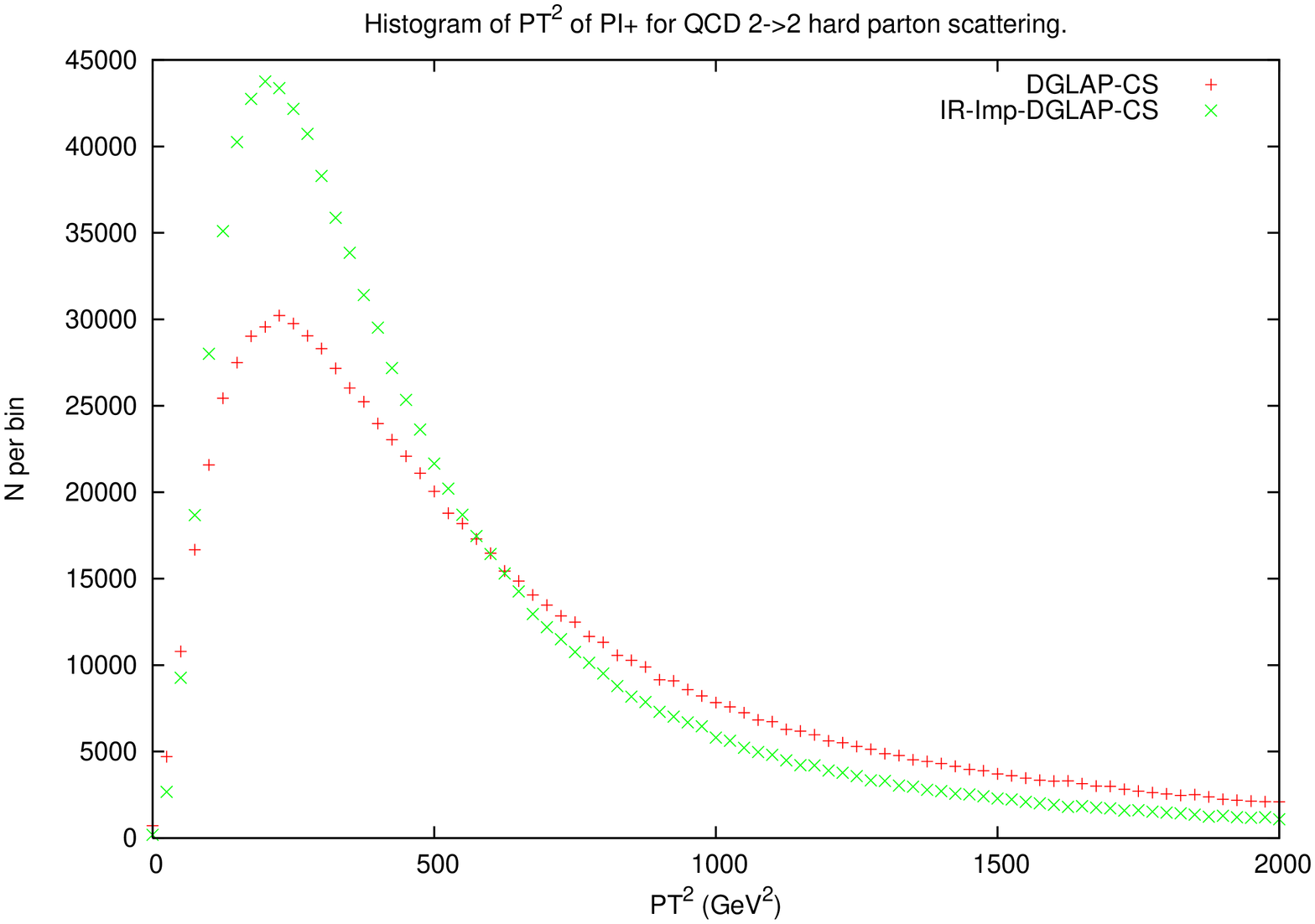,angle=270,
                                        width=50mm}}}
\end{picture}
\vspace{ -2mm}
\caption{\sf Distributions of the ISR parton energy fraction (a), $P_T^2$ 
(b), $\pi^+$ energy fraction (c), and $P_T^2$ (d) for the $2\rightarrow 2$ hard
subprocess in HERWIG-6.5, with (blue/green) and without (red) IR-improvement -- {\bf preliminary}. 
}
\label{fig1qcdichep08}
\end{figure*}

Firstly, we note that the connection to the higher order kernels in Refs.~\cite{high-ord-krnls} is done~\cite{irdglap1}. 
Secondly, in the NS case, we find that the n=2 moment
is modified by $\sim 5\%$ when evolved with (\ref{dglap19}) 
from $2$GeV to $100$GeV with $n_f=5$
and $\Lambda_{QCD}\cong .2GeV$, for illustration. Thirdly, 
we have been able to use
(\ref{subp15b}) to resolve violation~\cite{sac-no-go,cat1} 
of Bloch-Nordsieck cancellation in 
ISR at ${\cal O}(\alpha_s^2)$ for massive quarks~\cite{qmass-bw}. 
Fourthly, the threshold resummation implied by (\ref{subp15b}) for single Z
production at LHC shows a $0.3\%$ QED effect and agrees with known exact
results in QCD -- see Refs.~\cite{qced,baurall,exactqcd}. Fifthly, we have a new scheme~\cite{irdglap2} for precision LHC theory: in an obvious notation,
\begin{equation}
\sigma =\sum_{i,j}\int dx_1dx_2F_i(x_1)F_j(x_2)\hat\sigma(x_1x_2s)
       =\sum_{i,j}\int dx_1dx_2{F'}_i(x_1){F'}_j(x_2)\hat\sigma'(x_1x_2s),
\label{sigscheme}
\end{equation}
where the primed quantities are associated with (\ref{dglap19}) in the
standard QCD factorization calculus. Sixthly, we have~\cite{qced} an attendant
shower/ME matching scheme, wherein, for example, in combining (\ref{subp15b})
with Herwig~\cite{herwig}, Pythia~\cite{pythia}, MC@NLO~\cite{mcnlo}
or new shower MC's~\cite{skrzjad}, we may use either
$p_T$-matching
or shower-subtracted residuals $\{\hat{\tilde{\bar\beta}}_{n,m}(k_1,\ldots,k_n;k'_1,\ldots,k'_m)\}$ to create a paradigm without double
counting that can be systematically improved order-by order in
perturbation theory -- see Refs.~\cite{qced}. 
Finally, we show in 
Fig.~\ref{fig1qcdichep08} the effects of using the new kernels
in a new version of Herwig6.5, Herwig6.5-YFS, on 
both the shower and pion energy fraction
and $p_T$ spectra for $2\rightarrow 2$ hard sub-processes at LHC energies -- as expected, the IR-improved sprectra are softer.
We look forward to the further exploration and development of the results
presented herein.
\begin{acknowledgments}
%
One of us (B.F.L.W) acknowledges helpful discussions with Prof. Bryan Webber
and Prof. M. Seymour. B.F.L. Ward also thanks Prof. L. Alvarez-Gaume and Prof. W. Hollik for the support and kind hospitality of the CERN TH Division and of the Werner-Heisenberg Institut, MPI, Munich, respectively, while this work was in progress.

Work partly supported by US DOE grant DE-FG02-05ER41399 and 
by NATO Grant PST.CLG.980342.
\end{acknowledgments}

\end{document}